\documentclass[twocolumn,aps,prd]{revtex4-2}
\usepackage{graphicx}
\usepackage{bm}
\usepackage{relsize}
\usepackage{multirow}
\usepackage{xfrac}
\begin{document}
\newcommand*{\bi}{\bibitem}
\newcommand*{\ea}{\textit{et al.}}
\newcommand*{\eg}{\textit{e.g.}}
\newcommand*{\ie}{\textit{i.e.}}
\newcommand*{\plb}[3]{Phys.~Lett.~B \textbf{#1}, #2 (#3)}
\newcommand*{\phrc}[3]{Phys.~Rev.~C~\textbf{#1}, #2 (#3)}
\newcommand*{\phrd}[3]{Phys.~Rev.~D~\textbf{#1}, #2 (#3)}
\newcommand*{\phrl}[3]{Phys.~Rev.~Lett.~\textbf{#1}, #2 (#3)}
\newcommand*{\pr}[3]{Phys.~Rev.~\textbf{#1}, #2 (#3)} 
\newcommand*{\npb}[3]{Nucl.~Phys.~B \textbf{#1}, #2 (#3)} 
\newcommand*{\ptp}[3]{Prog. Theor. Phys. \textbf{#1}, #2 (#3)}
\newcommand*{\prpt}[3]{Phys. Rep. \textbf{#1}, #2 (#3)}
\newcommand*{\ra}{\rightarrow}
\newcommand*{\dd}{\mathrm D\bar\mathrm D}
\newcommand*{\dpdm}{\mathrm D^+\mathrm D^-}
\newcommand*{\dndn}{\mathrm D^0\bar\mathrm D^0}
\newcommand*{\dspdsm}{\mathrm{D}_s^+{\mathrm D}_s^-}
\newcommand*{\kpkm}{\mathrm{K}^+{\mathrm K}^-}
\newcommand*{\knkn}{\mathrm{K}^0\bar{\mathrm K}^0}
\newcommand*{\pds}{\psi(2\mathrm S)}
\newcommand*{\rf}[1]{(\ref{#1})}
\newcommand*{\be}{\begin{equation}}
\newcommand*{\ee}{\end{equation}}
\newcommand*{\bea}{\begin{eqnarray}}
\newcommand*{\eea}{\end{eqnarray}}
\newcommand*{\nl}{\nonumber \\}
\newcommand*{\die}{e^+e^-}
\newcommand*{\jj}{\mathrm i}
\newcommand*{\cndf}{\chi^2/\mathrm{NDF}}
\newcommand*{\ndf}{\mathrm{NDF}}
\newcommand*{\minuit}{\texttt{MINUIT}~}
\newcommand*{\e}[1]{{\mathrm e}^{#1}}
\newcommand*{\dek}[1]{\!\times\!10^{#1}}
\newcommand*{\rd}{\mathrm d}
\newcommand{\p}{$p$-value }
\title{
Subthreshold poles in electron‐positron annihilation. D\bm{$\bar{\mathrm D}$} 
final states.}
\author{Peter Lichard}
\affiliation{Institute of Physics and Research Centre for Computational 
Physics and Data Processing, Silesian University in Opava, 746 01 Opava, 
Czech Republic}

\begin{abstract}
We show that the $\psi(2\mathrm S)$ subthreshold pole influences the cross 
section of the electron-positron annihilation into the $\mathrm D^+\mathrm D^-$ 
and $\mathrm D^0\bar\mathrm D^0$ final states. 
We perform a fit to the merged BES \cite{bes2008} and BESIII \cite{besiii2024} 
data, providing the cross section for $e^+e^-$ annihilation into those final 
states. The statistical significance of the $\psi(2\mathrm S)$ as a 
subthreshold pole is 8.2$\sigma$. A fit assuming it and seven resonances 
excels with the fit quality characterized by $\chi^2/\mathrm{NDF}=240.4/284$, 
which means a $p$-value of 97\%. 
\end{abstract}
\maketitle

\section{Introduction}
This paper follows our effort to understand the role of the subthreshold
poles (SP in what follows) in the amplitude of the $\die$ annihilation to
hadrons that commenced in Ref. \cite{kaonium} by identifying SPs possibly 
representing the $\kpkm$ and $\knkn$ bound states and continued in Ref.  
\cite{dspdsm}, where the SP indicated the G(3900) as an excited state of
the $\dspdsm$ molecule. As in our previous papers, the essential tool we use
is the vector-meson dominance (VMD) hypothesis \cite{vmd}. 

VMD is a useful phenomenological concept 
based on the assumption that the interaction of the hadronic system with the 
electromagnetic field is mediated by truly neutral vector meson resonances 
of negative C-parity. It is frequently used to interpret the $\die$ 
annihilation data into hadrons. Experimentalists use resonances with masses 
above the considered reaction's threshold to fit the salient features of 
excitation curves (peaks, bumps, dips, steep slopes). VMD has also been used 
in modeling dilepton production in hadronic \cite{clp} and nuclear 
\cite{ruppert} collisions. 

Subthreshold poles are very well-known in hadronic reactions. 
A typical example is the nucleon pole in the pion-nucleon scattering. Even if 
it is not accessible in an experiment, it has a strong influence on the
cross section. 
Owing to that, its existence can be proven by analyzing the forward scattering 
amplitude obtained from phase-shift analyses, and its residue proportional 
to the $\pi$NN coupling constant can be determined with reasonable accuracy 
\cite{pin}. Guided by this analogy, we plan to continue investigating the 
role of subthreshold poles in the $\die$ annihilation into hadrons, which we
commenced in Ref. \cite{kaonium}.

To our knowledge, nobody has systematically investigated the analytic 
properties of the amplitudes of the $\die$ annihilation to various hadronic 
systems. Here, we will assume that they are similar to those of hadronic
amplitudes, discussed, \eg, in \cite{analprop}. 

From general principles of causality, locality of interactions, and unitarity, 
the amplitude is a real analytic function in the complex $s$-plane with the cut 
along the positive real axis running from the process threshold to infinity
(called the physical cut). 
The stable states, i.e., those which do not decay into the considered final 
state, are represented by poles lying in the real axis below the threshold. 
Resonances are represented by the pairs of complex conjugate poles situated 
on the higher Riemann sheets, accessible through the physical cut. 
The contribution of resonance to the amplitude on the upper branch of the 
physical cut is usually parametrized by some form of the Breit-Wigner 
formula. They run from the simplest one with constant mass and width to 
the most sophisticated with running ($s$-dependent) mass and variable width, 
which are related through the once-\cite{isgur} or twice-\cite{gs} 
subtracted dispersion relations. 

In this paper, we first show that an SP influences the electron-positron 
annihilation cross section into the $\dd$ final states with statistical
significance of 7.9$\sigma$. Its mass came out as $3691\pm56$~MeV. Taking
into account that the only charmonium with the mass lying in this
interval is the $\pds$, we fix the SP mass at 3686.097~MeV \cite{pdg2024}
and repeat the fit. Comparison with the fit without the SP yields the
$\pds$ pole significance of 8.2$\sigma$. Local significances show that
the $\pds$ pole is present equally in both $\die\ra\dpdm$ and $\die\ra\dndn$
processes.

\section{Model}
\label{model}
\begin{figure}[b]
\includegraphics[width=0.39\textwidth,height=0.13\textwidth]{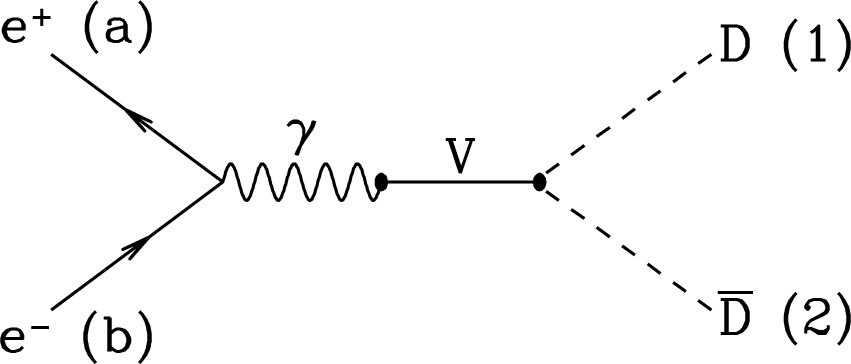}
\caption{\label{fig:ee2ddbar}Feynman diagram defining our model}
\end{figure}

For the description of the electron-positron annihilation into a $\dd$ pair
(D represents D$^+$ or D$^0$), we use a Vector Meson Dominance (VMD) model 
based on the Feynman diagram depicted in Fig.~\ref{fig:ee2ddbar} and the 
interaction Lagrangian
\be
\label{lagr}
{\cal
L}_{V\!\phi}(x)=\jj{g_{V\!\phi}}V_\mu(x)\left\{[\partial^\mu\phi^\dagger(x)]
\phi(x)-\phi^\dagger(x)\partial^\mu\phi(x)\right\},
\ee
where $V^\mu(x)$ denotes the (hermitian) vector field and $\phi(x)$ the 
pseudoscalar D field. The $\gamma V$ junction is parametrized as 
$eM_V^2/g_V$ in analogy with the $\gamma\rho^0$ junction 
$eM^2_{\rho^0}/g_\rho$. The formula for the cross section of the $\die$ 
annihilation into a $\dd$ pair based on the VMD model 
with $n$ resonances is derived in the Appendix. It sounds   
\be
\label{sigma}
\sigma(s)=\frac{\pi\alpha^2}{3s}\left(1-\frac{4m_{\mathrm D}^2}{s}\right)
^{\!\!{3/2}}\!
\left|\sum_{k=1}^n
\frac{\sqrt{Q_k}\,e^{\jj\delta_k}}{s-M_k^2+\jj M_k\Gamma_k}
\right|^2\!.
\ee
In this formula, $M_k$ and $\Gamma_k$ are the mass and width of the $k$th 
resonance, respectively, $Q_k=R_k^2$, where $R_k=M_k^2g_{V_k\phi}/g_{V_k}$, 
and $\delta_k$ is an additional phase shift. We will treat $M_k$s, 
$\Gamma_k$s, $Q_k$s, and $\delta_k$s as free parameters  (except $\delta_1$, 
which is kept at 0). We will determine their mean values and dispersions by 
fitting the formula \rf{sigma} to experimental cross sections using the 
standard $\chi^2$ criterion \cite{minuit}. 

\section{The explored data}

The experimentalists working at the spectrometer BESIII situated at 
the electron-positron collider BEPCII based at the IHEP laboratory
in Beijing, China, have recently published \cite{besiii2024} precise data 
on $\die$ annihilation to the $\dd$ final states. In the Supplemental
Material \cite{suppl}, they provide the Born cross sections of the
$\die\ra\dndn$ (Table 1) and $\die\ra\dpdm$ (Table 2) processes at 150 
cms energies between 3.80 and 4.95 GeV. Additionally, correction factors 
are tabulated, allowing for the conversion of the Born cross sections 
to the dressed ones.
The dressed cross sections were a subject of the fit presented in the 
Supplemental Material \cite{suppl} with results $\cndf=346/276$, which
implies the \p of 0.27\%. We will follow the 
experimentalists and use the dressed cross sections also in this work.

In the previous version of this study \cite{version2}, we used only the 
BESIII data \cite{besiii2024} and got evidence of the presence of the 
$\pds$ subthreshold pole in both $\die\ra\dndn$ and $\die\ra\dpdm$ processes. 
The anonymous referee has warned us that the data do not cover an important 
region between the threshold and 3.8~GeV, which contains the $\psi(3770)$ 
resonance. When searching for relevant data,  we found guidance in Ref.
\cite{ShamovTodyshev}, where a careful analysis of this region was done. 
We have chosen data from the BES Collaboration \cite{bes2008}, which were 
taken at the same IHEP laboratory, but at the BEPC collider with a 
conventional magnetic detector BES-2. Their Table 2 presents the cross 
section of both processes of interest at 14 energy points. We merged them 
with the BESIII data and will determine the free parameters of our model 
by fitting the formula \rf{sigma} to the complete set.

\section{A subthreshold pole appearing}

The BESIII Collaboration \cite{besiii2024} assumed in their fit that all 
parameters of the $\die\ra\dpdm$ and $\die\ra\dndn$ processes are the same.
Here, we will consider the differences between the two processes. 
Their origins are physical (different masses of $\mathrm D^+$ and 
$\mathrm D^0$ and possibly different couplings to intermediate vector
mesons), as well as experimental (different acceptances rooted in different 
methods of identifying particles in final states).
We will split each $R_k$ and $\delta_k$ parameter into two. Those pertinent 
to the $\dpdm$ final states will be marked by subscript $a$, those 
to the $\dndn$ ones by $b$. We will assume that the resonances' parameters 
$M_k$s and $\Gamma_k$ are the same for both processes.

To help the minimization code cope with many parameters, we started the
fitting with two resonances. We gradually added further, using the already
found parameters as the starting ones for the next run.

In this way, we reached the six-resonance fit with $\cndf=370.6/294$ and a 
\p of 0.16\%. To improve the fit, we added a new resonance with unconstrained
parameters. The quality of the fit increased to $\cndf=289.2/288$,
\p= 47\%. The statistical significance of the new resonance is 7.9$\sigma$,
estimated by utilizing the changes in likelihood values $\delta(-2\ln
L)=81.4$ and in the number of degrees of freedom $\delta(\ndf)=6$.

But the parameters of the new ``resonance'' are surprising: $M=(3691\pm56)$~MeV 
and $\Gamma=(0\pm160)$~MeV. They describe not a usual Breit-Wigner resonance, 
but an SP lying on the real axis of the complex $s$-plane
below the reaction threshold of 3729.68(10)~MeV (based on the D$^0$ mass
from \cite{pdg2024}).

When trying to identify the SP we have just discovered with an existing
state, we should not be fooled by its zero width. What looks like a stable 
state from the point of view of the $\dd$ final states may decay in other
processes. There are two obvious candidates with the $J^{PC}=1^{--}$ quantum 
numbers and the quark composition ensuring the strong coupling to the 
$D\bar D$ system: $J/\psi$(1S) with mass about 3097 MeV and $\pds$ with mass
3686.097(11) MeV \cite{pdg2024}. The mass of the SP we have found is by 
(5$\pm$57)~MeV higher than that of the $\pds$ and compatible with it. This 
fact leads us to choose the $\pds$ for further processing.

\section{Results}

We reperform a fit with six resonances and an SP with the mass fixed at that
of the $\pds$ and vanishing width. The goodness of the fit is characterized
by $\cndf=289.5/290$ and \p= 50\%. Comparison with the fit with six
resonances only implies the $\pds$ statistical significance as an SP of 
8.2$\sigma$ (based on $\delta(-2\ln L)=81.1$ and $\delta(\ndf)=4$).

To increase the fit quality further, we added another resonance, totaling
the number of resonances that accompany the $\pds$ SP to seven. This move has
resulted in $\cndf=240.4/284$ and \p= 97\%. The contributions of the $\dpdm$
and $\dndn$ channels to the $\chi^2$ are 119.5 and 120.9, respectively. 
Statistical significance against the previous solution, calculated from 
$\delta(-2\ln L)=49.1$ and $\delta(\ndf)=6$, is 5.7$\sigma$. 

We show the detailed results in Table \ref{tab:duo_parms}. In addition
to the parameters of the basic formula \rf{sigma}, we present also the local
significances of the SPs in the $\dpdm$ and $\dndn$ amplitudes and those
of resonances. We estimated them as $Q_k/\sigma_{Q_k}$.
\begin{table*}[b]
\caption{\label{tab:duo_parms}Parameters of the joint fit to the merged
$\dpdm$ (subscript $a$) and $\dndn$ (subscript $b$) data 
using Eq. \rf{sigma} with the $\pds$ SP and seven 
resonances. The masses $M$ and widths $\Gamma$ are assumed to be 
the same in both processes. Also, the local significances $S_a$ and $S_b$
are shown.} 
\begin{tabular}{lcccccccc}
\hline 
\multicolumn{8}{c}{$\cndf=240.4/284$~~~~~~~~~~~~~~~~~\p= 97\%}\\
\hline
$k$ & $1\,\equiv \pds$ & $2\,\equiv\psi(3770)$ & 3 & 4 & 5 & 6 & 7 & 8 \\
\hline
$M$ (MeV) &~3686.097(f) & 3779.13(76) & 3897$\pm$13 & 4026.6$\pm$1.9 &
 4101.7$\pm$6.3 & ~4207.8$\pm$3.3 & 4408.6$\pm$6.1 & 4572.8$\pm$6.8 \\
$\Gamma$ (MeV)&  0(f) & 27.1$\pm$1.6 & 149$\pm$16 & 29.4$\pm$3.1 &
94$\pm$20 & 41.7$\pm$5.4  & 110$\pm$12 & 46$\pm$14\\
$Q_a$ (GeV$^4$) & $52.2\pm6.6$ & 6.42(84) & 2.39(96) &0.134(35)&
0.16(11) & 0.0045(18)& 0.035(11) & 1.15(86)$\dek{-3}$  \\
$Q_b$ (GeV$^4$) & $53.7\pm6.3$ &5.81(76) & 2.06(79) & 0.131(34)&
0.139(94)& 0.0048(17) & 0.0261(81) & 0.96(64)$\dek{-3}$ \\
$\delta_a$ (rad)& 0(f)& $-2.757(93)$ & 2.44(25) & $-0.80(18)$&
1.95(23) &2.30(25) &1.21(13) & -0.88(30) \\
$\delta_b$ (rad)& 0(f)&$-2.729(88)$ & 2.44(27) & $-0.87(15)$ &
 1.98(21) & 2.20(25) & 0.98(13) & -1.44(31)\\
$S_a$ & 7.9$\sigma$ &7.6$\sigma$ &2.5$\sigma$ &3.8$\sigma$ &1.5$\sigma$ 
&2.5$\sigma$ &3.2$\sigma$ & 1.3$\sigma$ \\
$S_b$ & 8.5$\sigma$ &7.6$\sigma$ &2.6$\sigma$ &3.9$\sigma$ &1.5$\sigma$ 
&2.8$\sigma$ & 3.2$\sigma$ & 1.5$\sigma$ \\
\hline
\end{tabular}
\end{table*}

\begin{figure}[h]
\includegraphics[width=0.483\textwidth,height=0.34\textwidth]
{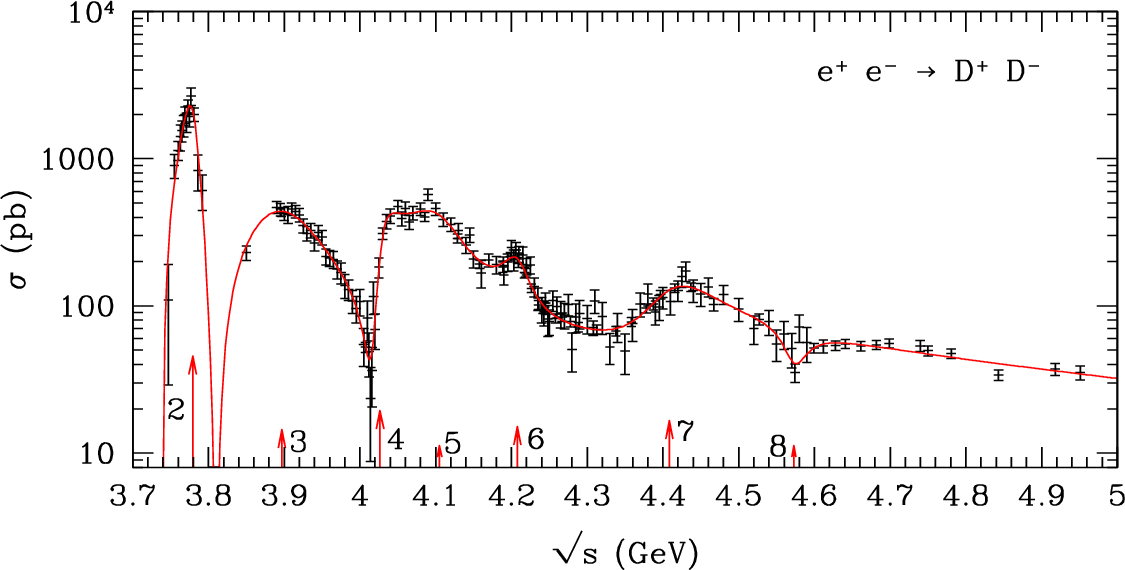}
\caption{\label{fig:DpDm}The $\die\ra\dpdm$ cross section as it follows
from the joint fit to both processes assuming the $\pds$ SP with seven 
resonances, summarized in Table \ref{tab:duo_parms}.}
\end{figure}

\begin{figure}[h]
\includegraphics[width=0.483\textwidth,height=0.34\textwidth]
{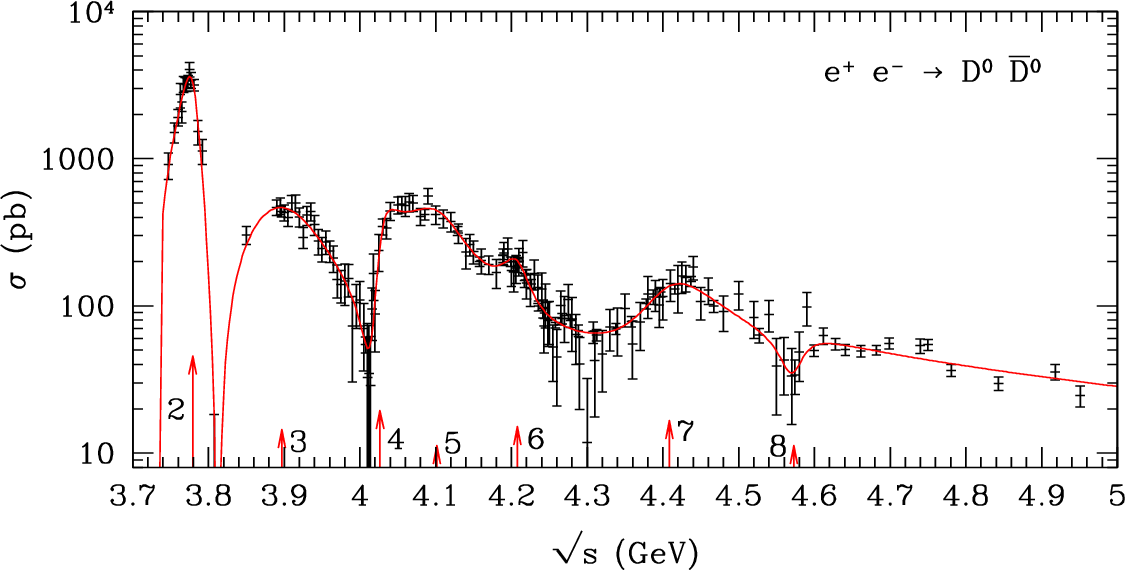}
\caption{\label{fig:D0aD0}The $\die\ra\dndn$ cross section as it follows
from the joint fit to both processes assuming the $\pds$ SP with seven 
resonances, summarized in Table \ref{tab:duo_parms}.}
\end{figure}

The $\die\ra\dpdm$ and $\die\ra\dndn$ cross sections calculated from the 
parameters shown in Table~\ref{tab:duo_parms} are compared to data in 
Figure~\ref{fig:DpDm} and Figure~\ref{fig:D0aD0}, respectively.

The main aim of this paper is to contribute to the effort to
understand the role of the SPs in the $\die$ annihilation into hadrons.
However, because the parameters of the resonances we found indicate that 
the unprecedentedly precise BESIII $\dd$ data may contribute significantly 
to the spectroscopy of the $c\bar c$ mesons, it is necessary to discuss 
this topic as well.

\begin{figure}[h]
\includegraphics[width=0.483\textwidth,height=0.31\textwidth]{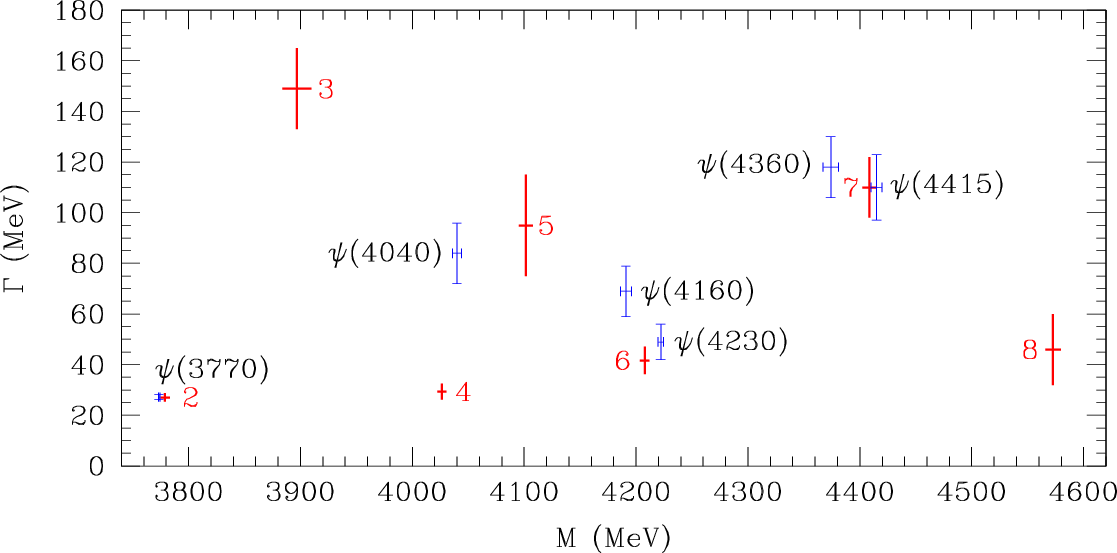}
\caption{\label{fig:mg_psi2S_7R}Comparison of the resonances found here
(marked with simple numerals) with those in the PDG 2024 tables, distinguished 
by short lines at endpoints.}
\end{figure}

It may be instructive to compare the results on resonances we obtained 
by fitting the very precise BESIII data \cite{besiii2024} supplemented
in the $\psi(3770)$ region by the BES data \cite{bes2008}. To facilitate the
discussion, in Fig. \ref{fig:mg_psi2S_7R} we also provide the $M-\Gamma$ 
plot containing the PDG 2024 \cite{pdg2024} $c\bar c$ resonances together 
with those resulting from
our analysis. The latter are supplied with their $k$-numbers from Table
\ref{tab:duo_parms}. The same are also used by arrows depicting the
resonance positions in Figs. \ref{fig:DpDm} and \ref{fig:D0aD0}. 

The mass 3779.13(76)~MeV and width 27.1$\pm$1.6~MeV of resonance \#2
agree well with the ``Our Average'' $\psi(3770)$ PDG 2024 values 
3778.1$\pm$0.7~MeV and 27.5$\pm$0.9~MeV, respectively.

The resonance \#3 with $M_3=3897\pm13$~MeV and $\Gamma_3=149\pm16$
corresponds to the G(3900) resonance, the nature of which is still debated. 
Our values are compatible with $M=3898.4\pm0.9$~MeV and 
$\Gamma=127.5\pm6.7$~MeV found in a recent global coupled-channel analysis 
\cite{satoshi}. No mention of G(3900) can be found in the PDG 2024 Tables.

The broad region around 4080 MeV is characterized by a steep leading 
edge and a more gentle decline at the end. This behavior cannot be 
described by a single resonance, \eg, $\psi(4040)$. Our fit showed that
the BESIII data requires two resonances instead, one narrow and one
wider. Their parameters are 
$M_4=4026.6\pm1.9$~MeV, $\Gamma_4=29.4\pm3.1$~MeV and 
$M_5=4101.7\pm6.3$~MeV, $\Gamma_5=94\pm20$~MeV. 

The resonances $\psi(4160)$ and $\psi(4230)$ have not appeared from our
fit. Instead of them, a resonance with parameters $M_6=4207.8\pm3.3$~MeV, 
$\Gamma_6=41.7\pm5.4$~MeV has. This is the same situation as with the
precise $\die\ra\dspdsm$ data \cite{besiii2024b} analysed in Ref. \cite{dspdsm}.
The ``extra'' resonance there has the parameters $M=4212.9\pm4.7$~MeV, 
$\Gamma=48.4\pm8.7$~MeV, in full conformity with our resonance \#6.

The PDG's $\psi(4360)$ did not show up in our fit. The reason may be its weak
affinity to the $\dd$ channel. We refer to a nonobservance of its $\dd$ decay,
citing the PDG 2024.

Our resonance \#7 perfectly agrees with the PDG's $\psi(4415)$. Just the mass
is smaller, $4408.6\pm6.1$ against $4415\pm5$~MeV.

Resonance \#8 with $M_8=4572.8\pm6.8$~MeV, $\Gamma_8=46\pm13$~MeV manifests 
itself as a dip in both excitation curves. However, its local significance is 
minor: 1.3$\sigma$ in the $\dpdm$ amplitude and 1.5$\sigma$ in the $\dpdm$ one.

\section{Conclusions}
The inclusion of an SP significantly improves the quality of a
six-resonance fit to the combined BES \cite{bes2008} and BESIII
\cite{besiii2024} data, expressed by the $p$-value rising from 
0.16\% to 47\% and the SP statistical significance of 7.9$\sigma$. 
The obtained SP position is very close to the $\pds$ resonance mass, which has 
led us to hypothesize that the $\pds$ state, a resonance in other processes, 
behaves in the $\die\ra\dd$ process like a subthreshold pole with a 
statistical significance of 8.2$\sigma$. 

Finally, we have supplemented the $\pds$ SP with seven resonances and
performed a fit. It resulted into $\cndf=240.4/284$ and \p= 97\%. 
 
Our study suggests that it may be impossible to get a good fit to the
$\die\ra\dd$ process without considering a subthreshold pole. More
generally, we follow up with our other studies \cite{kaonium,dspdsm}
showing the importance of the often-overlooked role of subthreshold poles,
which can significantly affect some processes.


\acknowledgements
The author thanks the anonymous referee for showing him importance of the 
$\psi(3770)$ region, Josef Jur\'{a}\v{n} for enlightening him on some statistics
aspects and for valuable remarks on the text, and Veronika Gintnerov\'{a} 
for checking the transfer of information from the computer outputs to the 
manuscript. 

\appendix*
\section{Derivation of formula \rf{sigma}}
The Lagrangian \rf{lagr} implies for the V$\dd$ vertex in Fig.
\ref{fig:ee2ddbar} the expression $g_{V\phi}(p_1-p_2)^\rho$. Together with
the $eM_V^2/{g_V}$ for the $\gamma V$ junction and standard Feynman rules,
it leads to the following amplitude of the $\die\ra\dd$ process 
($p=p_a+p_b$, $s=p^2$, electron mass neglected): 
\bea
\jj{\cal M}_{fi}&=&\jj e\bar{v}(p_a,s_a)\gamma_\mu u(p_b,s_b)\,
\frac{-\jj g^{\mu\nu}}{s}\,\frac{eM_V^2}{g_V}\nl
&\times&\jj\frac{-g_{\nu\rho}+\frac{p_\nu p_\rho}{M_V^2}}{s-M_V^2+\jj M_V
\Gamma_V}\,g_{V\phi}(p_1-p_2)^\rho \ .\nonumber
\eea 
After some editing and introducing the notation $R_V=M_V^2g_{V\phi}/g_V$, the 
amplitude squared takes the form 
\bea
|{\cal M}_{fi}|^2&=&\frac{e^4}{s^2}\left|\frac{R_V}{s-M_V^2+\jj M_V
\Gamma_V}\right|^2\!(p_1-p_2)^\mu(p_1-p_2)^\nu\nl
&\times&{\bar v}(p_a,s_a)\gamma_\mu u(p_b,s_b){\bar u}(p_b,s_b)\gamma_\nu
v(p_a,s_a)\ .  \nonumber
\eea
Averaging over initial spin states leads to [$t=(p_1-p_a)^2$] 
\bea
\overline{|{\cal M}_{fi}|^2}&=&\frac{2e^4}{s^2}\left|\frac{R_V}
{s-M_V^2+\jj M_V\Gamma_V}\right|^2\nl
&\times&(-st-t^2+2m_{\rm D}^2t-m_{\rm D}^4)\ .\nonumber
\eea
Inserting this into the differential cross-section formula 
\[
\frac{\rd\sigma}{\rd t}=\frac{1}{64\pi s}\frac{1}{|\vec p_a|^2}
\overline{|{\cal M}_{fi}|^2}
\]
and integrating over $t$ from $t_1=m_{\rm D}^2-\frac{1}{2}(1+v_{\rm D})$
to $t_2=m_{\rm D}^2-\frac{1}{2}(1-v_{\rm D})$, where $v_{\rm D}$ is the
D meson speed, and putting $R_V=\sqrt{Q_V}$, we finally get 
\be
\label{sigma1V}
\sigma(s)=\frac{\pi\alpha^2}{3s}\left(1-\frac{4m_{\mathrm D}^2}{s}\right)
^{\!\!{3/2}}\!
\left|
\frac{\sqrt{Q_V}}{s-M_V^2+\jj M_V\Gamma_V}
\right|^2\!.
\ee
If the intermediate vector meson is a resonance, we can obtain in the same
way the relations below. 
\be
\label{vdd}
\Gamma_{V\ra\dd}=\frac{M_Vg_{V\phi}^2}{48\pi}\left(1-\frac{4m_{\rm D}^2}
{M_V^2}\right)^{3/2}
\ee
and
\be
\label{vee}
\Gamma_{V\ra\die}=\frac{4\pi\alpha^2}{3}\frac{M_V}{g_V^2}\ .
\ee
Combining Eqs. \rf{sigma1V}, \rf{vdd}, and \rf{vee} we can directly relate 
the parameter $Q_V$ to observable quantities 
\be
\label{qv}
Q_V=\frac{36M_V^2}{\alpha^2}\ \frac{\Gamma_{V\ra\dd}\,\Gamma_{V\ra\die}}
{\left(1-\frac{4m_{\rm D}^2}{M_V^2}\right)^{3/2}}\ .
\ee
The relation \rf{qv} is valid only for resonances, whereas 
\rf{sigma1V} is valid also for SP (with $\Gamma_V=0$).

The relation \rf{qv} is one of the reasons why we prefer $Q_V$ over $R_V$. 
Another is that the $Q_V$ and its dispersion can provide information about the 
local significance of the corresponding resonance or SP. 

Generalizing Eq. \rf{sigma1V} to the case of $n$ interferring resonances 
leads to Eq. \rf{sigma}.

\end{document}